\documentclass{elsarticle}
\usepackage[margin=3cm,top=2cm]{geometry}
\usepackage[usenames]{color}
\usepackage{graphicx}
\usepackage{caption}
\usepackage{subcaption}
\usepackage{float}
\usepackage{amsmath}
\usepackage{amsfonts}
\usepackage{epstopdf}
\usepackage{amsthm}
\usepackage{acronym}

\definecolor{myGreen}{rgb}{0.17254902, 0.62745098, 0.17254902}
\definecolor{myPurple}{rgb}{0.5, 0.0, 0.5}

\renewcommand{\vec}[1]{\mathbf{#1}}
\newcommand{\mat}[1]{\mathsf{#1}}
\def\mb{\mathbf}

\begin{document}

\begin{frontmatter}

\title{Data-driven molecular modeling with the generalized Langevin equation}
\date{December 1, 2019}
\author[1]{Francesca Grogan\corref{cor1}}
\ead{francesca.grogan@pnnl.gov}
\author[2,3]{Huan Lei}
\ead{leihuan@msu.edu}
\author[4]{Xiantao Li}
\ead{xli@math.psu.edu}
\author[1,5]{Nathan A.~Baker}
\ead{nathan.baker@pnnl.gov}
\address[1]{ Pacific Northwest National Laboratory, Richland, WA 99352, United States}
\address[2]{ Department of Computational Mathematics, Science and Engineering, Michigan State University, East Lansing, MI 48824, United States}
\address[3]{ Department of Statistics and Probability, Michigan State University, East Lansing, MI 48824, United States}
\address[4]{ Department of Mathematics, Pennsylvania State University, State College, PA 16801, United States}
\address[5]{ Division of Applied Mathematics, Brown University, Providence, RI 02912, United States}
\cortext[cor1]{Corresponding author}

\begin{abstract}
The complexity of \acl*{MD} simulations necessitates dimension reduction and coarse-graining techniques to enable tractable computation.
The \acf*{GLE} describes coarse-grained dynamics in reduced dimensions.
In spite of playing a crucial role in non-equilibrium dynamics, the memory kernel of the \acs*{GLE} is often ignored because it is difficult to characterize and expensive to solve.
To address these issues, we construct a data-driven rational approximation to the \acs*{GLE}.
Building upon previous work leveraging the \acs*{GLE} to simulate simple systems,  we extend these results to more complex molecules, whose many degrees of freedom and complicated dynamics require approximation methods.
We demonstrate the effectiveness of our approximation by testing it against exact methods and comparing observables such as autocorrelation and transition rates.
\end{abstract}


\end{frontmatter}

\section{Introduction}
Molecular dynamics methods simulate atomic trajectories using Newton's second law of motion.
Full atomic-detail \ac{MD} simulations are often prohibitively expensive due to the complexity and size of the systems under study. 
Model reduction based on surrogates~\cite{Gubskaya2007,Nance2015,Lei_Yang_MMS_2015, Razi2018,Lei_Li_Gao_2019,Zhao2019} and projection operators~\cite{Mori1965,Zwanzig1973} is a popular approach for reducing dimension and complexity in a wide range of computational science applications.
One such model is the \ac{GLE}~\cite{Mori1965,Zwanzig1973}, which describes the system in terms of collective degrees of freedom and simulates dynamics in terms of coarse-grained \acp{CV}.
The \ac{GLE} reduces the problem size by only explicitly representing the dynamics of these \acp{CV}; the remaining degrees of freedom are described implicitly.
\ac{GLE}-based approaches have been successfully used in a variety of application areas~\cite{Adelman2007, Turq1977, Cordoba2012, Demery2014,  Wu2018, ariel2007testing}.
An important component of the \ac{GLE} is a time-dependent memory kernel that accounts for the implicit degrees of freedom and their impact on the evolution of the explicitly resolved \acp{CV}.
This memory term plays a crucial role in non-equilibrium dynamics but is often hard to characterize and evaluate, particularly for high-dimensional systems~\cite{Darve2009, Li2010, Jung_Hanke_JCTC_2017}.
The kernel is sometimes simplified to reduce computational requirements; however, this often renders the model unable to accurately represent system dynamics~\cite{Guardia1985, Lei2010, Shugard1977}.

Ideally, construction of the memory kernel should balance computational cost and accuracy.
For theoretical convergence analysis and error bounds for various memory kernel approximations, see~\cite{Zhu2018}.
Previous work by the authors~\cite{Lei2016} introduced a data-driven approach to parameterize the \ac{GLE} memory kernel via a rational approximation in Laplace space.
This modeling ansatz---along with the introduction of appropriate auxiliary variables---transforms the \ac{GLE} into an extended system driven by white noise, where the second \ac{FDT}~\cite{Kubo1966} can be satisfied by properly choosing the covariance matrix.
Numerical studies on simple systems (a tagged particle in solvent) show that this approach can successfully characterize the non-equilibrium dynamics beyond Einstein's Brownian motion theory and accurately predicts observables such as transition rates between a double-well potential.
In the data-driven algorithm, modeling accuracy relies on the approximation order of the memory kernel.
Data-driven model reduction methods have also been developed by others using a variety of approaches~\cite{lin2019data, lu2016comparison, ma2018model, russo2019deep}.
Additionally, Zhu and Venturi have demonstrated the use of polynomial approximations of memory kernels for \ac{GLE}-like problems~\cite{zhu2018faber} as well as a first-principles method for systems with local interactions~\cite{Zhu2020}.

In this work, we extend the data-driven parameterization approach~\cite{Lei2016} to construct a reduced model for the small molecule system of \ac{BnBr} in explicit water.
We recently developed a data-driven approach~\cite{Lei_Li_Gao_2019} for uncertainty quantification of the equilibrium properties (e.g., solvation energy) with
respect to the non-Gaussian conformation fluctuations using this solvated \ac{BnBr} system.
To quantify the non-equilibrium dynamics, the non-Markovian memory will need to be accurately constructed.
In particular, this system is more complex than the benchmark problems considered previously by us~\cite{Lei2016}: its energy landscape has multiple energetic minima and 
both the intra- as well as inter-molecular interactions contribute to the energy-dissipation process.
On the other hand, the small size of \ac{BnBr} allows \ac{MD} simulations to achieve near-ergodic sampling of its conformational space within a tractable amount of time.
The transition rate between the two conformational states can be directly evaluated by \ac{MD} simulation and compared with the predictions from the reduced model.
Recently, similar work has been reported by Lee and co-workers~\cite{Lee2019}, where a reduced model of the molecule alanine dipeptide is constructed by the \ac{GLE} in terms of two dihedral angles and is then parameterized through time-series expansions.
In the current study, we present an alternative approach that constructs the memory kernel in Laplace space based on a modification of our earlier approach.
These modifications were made to accommodate the more complex gradient system we study in this work, see Section~\ref{sec:memory_approx} for details.
An advantage of our approach is that accuracy can be adaptively tuned by adjusting the order of the memory kernel approximation.
We demonstrate the applicability of our \ac{GLE} method on model reduction for molecules in aqueous environments.

The paper is organized as follows.  
Section \ref{s:methods} introduces the \ac{GLE} and presents our methodology for constructing the data-driven reduced-order model.  
We discuss our simulation setup at the end of Section \ref{s:methods}.
In Section \ref{s:results}, we present results testing our approximation against exact methods and comparing observables such as autocorrelation and transition rates which show the exact memory term is well-modeled by its data-driven parametrization.
We briefly conclude and discuss avenues for future work in Section \ref{s:conclusion}.

\section{Methods}\label{s:methods}

We begin in Section \ref{sec:preliminary} by introducing the \ac{GLE} and the \acp{CV} used.
Section \ref{sec:memory_approx} discusses two approaches to build a rational approximation to the \ac{GLE} memory kernel.
Using extended dynamics to represent the \ac{GLE} is presented in Section \ref{sec:ex_dynamics}, with initial and noise conditions detailed in Section \ref{sec:fdt}.  
Finally, we provide simulation setup details in Section \ref{sec:sim_setup}.

\subsection{Preliminaries}\label{sec:preliminary}

Before introducing the \ac{GLE}, we discuss the \acp{CV} to be calculated from our \ac{BnBr} simulations.
We perform \ac{PCA} on the \ac{BnBr} atom positions \(\vec{x}(t): [0, \infty) \mapsto \mathbb{R}^{N}\) and velocities \(\dot{\vec{x}}(t)\) obtained from an \ac{MD} trajectory, where $N$ is the number of degrees of freedom in the system (usually $N = 3n - 6$ for $n$ atoms).
The covariance matrix $\mat{C} \in \mathbb{R}^{N \times N}$ is defined as
\begin{equation}
    \mat{C} = \left\langle \left( \vec{x}(t) - \left\langle \vec{x}(t) \right\rangle \right)
    \left( \vec{x}(t) - \left\langle \vec{x}(t) \right\rangle \right)^T \right\rangle,
    \nonumber
\end{equation}
where $\langle \cdot \rangle$ denotes the ensemble average with respect to the equilibrium distribution of $\vec{x}$.

We project the \ac{BnBr} trajectory onto the principal modes using the eigen-decomposition $\mat{C} = \mat{V} \mat{D} \mat{V}^T$ to obtain the principal components $\vec{q}(t): [0, \infty) \mapsto \mathbb{R}^{N}$
\begin{equation}
    \vec{q}(t) = \mat{V}^T \left(\vec{x}(t) - \left\langle \vec{x}(t) \right\rangle \right)
    \nonumber
\end{equation}
and associated velocities $\dot{\vec{q}}(t)$.
These principal components provide an understanding of the dynamic behavior of \ac{BnBr} by highlighting the dominant motions of the molecule.
In our study, we use the first principal component as our \ac{CV}; however, it is possible to generalize our method to multi-dimensional as well as nonlinear \acp{CV}.

We next introduce the \ac{CV} mass matrix to be used in the \ac{GLE}.
Generally, for a nonlinear \ac{CV}, $f(\vec{x})$, the mass matrix $\mat{M}$ is the diagonal matrix whose elements $M_{ii}$ are given by
\begin{equation}
    M_{ii} = {\left( \int \sum_{i=1}^n \frac{1}{\mu_i}{\left(\frac{\partial f}{\partial x_i} \right)}^2 \rho(\vec{x})d\vec{x} \right)}^{-1},
    \nonumber
\end{equation}
where $\mu_i$ is the mass associated with the $i$-th atom and $\rho$ is the {equilibrium} \ac{PDF}.
In the case of our linear \ac{CV}, $\mat{M}$ can simply be defined using the equipartition theorem:
\begin{equation}
    \mat{M} \left\langle \dot{\vec{ q}}\dot{\vec{ q}}^T\right\rangle = \beta^{-1} \mat{I},
    \nonumber
\end{equation}
where $\beta = \left( k_B T \right)^{-1}$, $k_B$ is the Boltzmann constant, $T$ is the temperature, and $\mat{I}$ is the $N$-dimensional identity matrix.
For details regarding the equivalence of the two formul\ae, see Lange and Grubm\"{u}ller~\cite{Lange2006}.
We note that the mass matrix may also be approximated as a function of the \ac{CV}, so that the \ac{GLE} consists of only coarse-grained terms~\cite{Lee2019}.

Given a mass matrix $\mat{M}$, the momentum $\vec{p}(t): [0, \infty) \mapsto \mathbb{R}^N$ is defined as 
\begin{align}
	\vec{p}(t) = \mat{M} \dot{\vec{q}}(t)
	\label{eq:momentum}
\end{align}
and the \ac{GLE} can be written as 
\begin{align} 
    \dot{\vec{p}} &= \vec{F}(\vec{q}) - \int_0^t \mat{K}(t-\tau) \dot{\vec{q}}(\tau) d\tau + \vec{R}(t),
    \label{e:md}
\end{align}
where $\vec{F(q)}: [0, \infty) \mapsto \mathbb{R}^N$ is the conservative force, $\mat{K}(t): [0, \infty) \mapsto \mathbb{R}^N \times \mathbb{R}^N$ is the time-dependent memory kernel function, and $\vec{R}(t): [0, \infty) \mapsto \mathbb{R}^N$ is the random noise modeled as a stationary Gaussian process with zero mean that satisfies the second \ac{FDT}:
\begin{equation}
    \left\langle \vec{R}(t) \vec{R}(t')^T \right\rangle = \beta^{-1}\mat{K}(t-t').
    \nonumber \\
\end{equation}

\subsection{Constructing a rational approximation to the memory kernel} 
\label{sec:memory_approx}

We define the correlation matrices $\mat{G}(t): [0, \infty) \mapsto \mathbb{R}^N \times \mathbb{R}^N$ and $\mat{H}(t): [0, \infty) \mapsto \mathbb{R}^N \times \mathbb{R}^N$ as
\begin{equation}
    \begin{aligned}
        \mat{G}(t) &= \left\langle \dot{\vec{p}}(t) \vec{q}(0)^T - \vec{F}(\vec{q}(t)) \vec{q}(0)^T \right\rangle \\
        \mat{H}(t) &= \left\langle \dot{\vec{q}}(t) \vec{q}(0)^T \right\rangle.
    \end{aligned}
    \label{eqn:gh}
\end{equation}
Right-multiplying by $\vec{q}(0)^T$, the \ac{GLE} (Eq.~\eqref{e:md}) becomes
\begin{equation}
    \mat{G}(t) = -\int_0^t \mat{K}(t - \tau) \mat{H}(\tau) d \tau
    \label{eq:GLE2}
\end{equation}
with the assumption $\left\langle \vec{R}(t) \vec{q}(0)^T \right\rangle = \mat{0}$; see~\cite{Chen2014} for details.

With $\mat{G}(t)$ and $\mat{H}(t)$ defined by Eq.~\eqref{eqn:gh}, we can solve Eq.~\eqref{eq:GLE2} by transferring this integral equation into frequency space using the Laplace transform~\cite{Linz1969}:
\begin{equation}
    \hat{\mat{G}}(\lambda) = \int_0^\infty \mat{G}(t)e^{-t/\lambda} dt,  \hspace{1em}
    \hat{\mat{H}}(\lambda) = \int_0^\infty \mat{H}(t)e^{-t/\lambda} dt,  \hspace{1em}
    \hat{\mat{K}}(\lambda) = \int_0^\infty \mat{K}(t)e^{-t/\lambda} dt,
    \label{e:laplace_eq}
\end{equation}  
such that Eq.~\eqref{eq:GLE2} becomes
\begin{equation}
    \hat{\mat{G}}(\lambda) = - \hat{\mat{K}}(\lambda) \hat{\mat{H}}(\lambda).
    \label{e:K}
\end{equation}
Taking $\lambda \rightarrow \infty$ of Eq.~\eqref{e:laplace_eq} gives
\begin{equation}
    \hat{\mat{G}}(\infty) = \int_0^\infty \mat{G}(t) dt, \hspace{1em}
    \hat{\mat{H}}(\infty) = \int_0^\infty \mat{H}(t) dt, \hspace{1em}
    \hat{\mat{K}}(\infty) = \int_0^\infty \mat{K}(t) dt.
    \label{e:laplace_eq_inf}
\end{equation}  
We note that the definitions of $\mat{G}$ and $\mat{H}$ differ from our previous work~\cite{Lei2016}, where $\mat{H}$ was 
defined as the velocity correlation matrix, i.e., $\langle \dot{\vec{q}}(t) \dot{\vec{q}}(0)^T\rangle$.
However, our previous choice led to numerical instability in the 
construction of $\hat{\mat{K}}(\lambda)$. In particular, the Markovian limit condition requires
$\lim_{\lambda \to \infty} \hat{\mat{G}}(\lambda) = - \lim_{\lambda \to \infty} \hat{\mat{K}}(\lambda)\hat{\mat{H}}(\lambda)$.
If we choose $\mat{H}(t) = \left\langle \dot{\mb q}(t) \dot{\mb q}(0)^T\right\rangle$, we need to evaluate the term 
$\lim_{\lambda \to \infty} \hat{\mat{H}}(\lambda) = \int_{0}^{+\infty} \left\langle \dot{\mb q}(t) \dot{\mb q}(0)^T\right\rangle dt$.
For the gradient system considered in the present work, we note that  
\begin{equation}
\begin{split}
\lim_{t \to \infty} \left\langle {\mb q}(t) \dot{\mb q}(0)^T\right\rangle &= \lim_{t \to \infty}  
\int \rho(\mb q(t) = \mb q'\vert \dot{\mb q}(0) = \mb v_0) \rho_{0}^v(\mb v_0) \mb q' \mb v_0^T d \mb v_0 \\
&= \int \rho_{\rm eq}^q(\mb q') \rho_{0}^v(\mb v_0) \mb q' \mb v_0^T d \mb v_0 \\
&\propto \int e^{-\beta U(\mb q')} e^{-\beta \mb v_0^T\mat{M}^{-1}\mb v_0/2} \mb q' \mb v_0^T d \mb v_0 \equiv 0
\end{split}
\nonumber
\end{equation}
where $\rho(\mb q(t) = \mb q'\vert \dot{\mb q}(0)= \mb v_0)$ represents the conditional probability of $\mb q(t)$ with the initial
condition $\dot{\mb q}(0) = \mb v_0$ and we take $\rho^{v}_0(\mb v_0)$, the probability density function of $\mb v_0$,  
to be the equilibrium density. 
Accordingly, $\lim_{\lambda \to \infty} \hat{\mat{H}}(\lambda) = 0$, yielding the ill-conditioning of the Markovian limit of 
Eq.~\eqref{e:K}. Using $\mat{H}(t)$ as defined in Eq.~\eqref{eqn:gh} does not result in this ill-conditioning.
(We note that for the numerical cases considered in \cite{Lei2016}, the dynamic equation does not contain 
the term $U(\mb q)$ and we do not encounter such difficulty).

With $\hat{\mat{G}}(\lambda)$ and $\hat{\mat{H}}(\lambda)$ sampled from \ac{MD} simulations, we 
construct the memory kernel $\hat{\mat{K}}(\lambda)$ in the form of 
\begin{equation}
    \hat{\mat{K}}(\lambda) \approx \left( \mat{I} - \sum_{m=1}^M \mat{B}_{m} \lambda^m \right)^{-1} \left( \sum_{m=1}^M \mat{A}_{m} \lambda^m \right),
    \label{e:ratK}
\end{equation}
where the terms of the expression are matrices $\mat{A}_m, \mat{B}_m \in \mathbb{R}^{N \times N}$.
The highest-order coefficients of an $M$-order expansion can be found through the limit of Eq.~\eqref{e:ratK}:
\begin{equation}
    \lim_{\lambda \rightarrow \infty} \hat{\mat{K}}(\lambda) = - \mat{B}_M^{-1} \mat{A}_M,
    \label{e:lim_k}
\end{equation}
as $\hat{\mat{K}}(\infty) = - \hat{\mat{G}}(\infty) \hat{\mat{H}}(\infty)^{-1} 
= - \left( \int_0^\infty \mat{G}(t)dt \right) \left( \int_0^\infty \mat{H}(t)dt \right)^{-1}$ by taking $\lambda \rightarrow \infty$. 
Note that $\hat{\mat{K}}(\infty)$ recovers the friction tensor in Markovian limit, i.e., the Markovian approximation is the zeroth-order GLE approximation,
\begin{equation}
	\dot{\vec{p}} = \vec{F}(\vec{q}) -\hat{\mat{K}}(\infty)\dot{\vec q}(t) + \vec{R}(t), 
\end{equation} 
where $\hat{\mat{K}}(\infty)$ is the friction tensor and is proportional to the diffusion tensor $\mat{D}$ by the Einstein relation $\hat{\mat{K}}(\infty) = k_B T \mat{D}^{-1}$.

Eq.~\eqref{e:lim_k} allows us to solve for either $\mat{A}_M$ or $\mat{B}_M$. 
To solve for the remaining $M-1$ unknown coefficients, there are two approaches.  
If high-order derivatives of $\mat{H}(t)$ and $\mat{G}(t)$ are available from the data at $t = 0$, then $\hat{\mat{K}}(\lambda)$ can be (semi-analytically) 
constructed by the first approach described below.
Alternatively, as the numerical evaluation of higher-order terms may introduce significant numerical error in the coefficient calculations, $\hat{\mat{K}}(\lambda)$ can be numerically constructed by a regression approach using $\hat{\mat{G}}(\lambda)$ and $\hat{\mat{H}}(\lambda)$ at interpolation points.

\paragraph{Approach 1}
The first approach involves coefficient matching and differentiation.
First, we perform a Taylor expansion of $\hat{\mat{K}}(\lambda)$
\begin{equation}
    \hat{\mat{K}}(\lambda) = \sum_{n=1}^\infty \frac{\hat{\mat{K}}^{(n)}(0)}{n!} \lambda^n.
    \label{e:k_taylor}
\end{equation}
Substituting this expression into the left-hand side of Eq.~\eqref{e:ratK} and matching with respect to $\lambda$, we obtain the formula
\begin{equation}
	\frac{\hat{\mat{K}}^{(n)}(0)}{n!} 
	= \mat{A}_n + \sum_{l+m=n} \mat{B}_l \frac{\hat{\mat{K}}^{(m)}(0)}{m!} .
	\label{e:match_taylor}
\end{equation}
We can then determine the terms $\hat{\mat{K}}^{(i)}$ by differentiating Eq.~\eqref{e:K}.
As an example, we compute the first-order coefficients.
In this case, we use Eq.~\eqref{e:match_taylor} to match coefficients with respect to $\lambda^1$:
\begin{equation}
	\hat{\mat{K}}^{(1)}(0) = \mat{A}_1,
\end{equation}
where we have used the fact that $\hat{\mat{K}}^{(0)}(0) = 0$.
To find an expression for the derivative $\hat{\mat{K}}^{(1)}(0)$, we differentiate Eq.~\eqref{e:K} and let $\lambda \rightarrow 0$
\begin{eqnarray}
	\hat{\mat{G}}^{(3)}(0) &=& - [\hat{\mat{K}}(0) \hat{\mat{H}}^{(3)}(0)
	+ 3 \hat{\mat{K}}^{(1)}(0) \hat{\mat{H}}^{(2)}(0)
	+ 3 \hat{\mat{K}}^{(2)}(0) \hat{\mat{H}}^{(1)}(0)
	+  \hat{\mat{K}}^{(3)}(0) \hat{\mat{H}}(0)] \nonumber \\
	&=& -3 \hat{\mat{K}}^{(1)}(0) \hat{\mat{H}}^{(2)}(0),
	\label{eq:diff_g}
\end{eqnarray}
noting that $\lim_{\lambda \rightarrow 0} \hat{\mat{K}}(\lambda) = \lim_{\lambda \rightarrow 0} \hat{\mat{H}}(\lambda) = \lim_{\lambda \rightarrow 0} \hat{\mat{H}}^{(1)}(\lambda) = 0$ since $\vec{q}$ and $\dot{\vec{q}}$ are uncorrelated. 
Integrating Eq.~\eqref{e:laplace_eq} by parts and letting $\lambda \rightarrow 0$ gives~\cite{Lei2016}
\begin{equation}
    \hat{\mat{G}}^{(i)}(0) = i! \cdot \mat{G}^{(i-1)}(0), \hspace{1em}
    \hat{\mat{H}}^{(i)}(0) = i! \cdot \mat{H}^{(i-1)}(0), \hspace{1em}
    \hat{\mat{K}}^{(i)}(0) = i! \cdot \mat{K}^{(i-1)}(0).
    \label{eq:ibp}
\end{equation}
Combining Eq.~\eqref{eq:ibp} with Eq.~\eqref{eq:diff_g}, we arrive at the following expressions for the first-order coefficients:
\begin{eqnarray}
	\mat{A}_1 &=& - \mat{G}^{(2)}(0) [\mat{H}^{(1)}(0)]^{-1} \\
	\mat{B}_1 &=& - \mat{A}_1 \hat{\mat{K}}(\infty)^{-1}.
\end{eqnarray}

\paragraph{Approach 2}
The second approach to solve for unknown coefficients also starts with Eq.~\eqref{e:lim_k}.
However, in this approach, the memory kernel is constructed using regression at discrete values of $\lambda$.  
For an $M$-order approximation, we choose a set of points $\lambda_1, \lambda_2, \ldots, \lambda_{2M-1} \in (0,\infty)$ and solve for the coefficients such that the approximate memory kernel (Eq.~\eqref{e:ratK}) interpolates the exact memory kernel at the chosen set of points. 
This results in $M-1$ nonlinear equations.  
With Eq.~\eqref{e:lim_k}, these equations comprise a nonlinear system of $M$ equations to be solved, which we can express as
\begin{equation}
	\vec{F}(\lambda_1, \cdots, \lambda_{2M-1}; \mat{A}_1, \cdots, \mat{A}_M,  \mat{B}_1, \cdots, \mat{B}_M) = 0. 
	\label{eq:app2}
\end{equation}
Continuing with our first-order example, this approach would result in the following $\vec{F}$: 
\begin{equation}
	\vec{F}(\lambda_1; \mat{A}_1, \mat{B}_1) = \left\{
                \begin{array}{ll}
                  (\mat{I} - \mat{B}_1\lambda_1)^{-1}(\mat{A}_1\lambda_1) -  \hat{\mat{K}}(\lambda_1)\\
                   \mat{B}_1 + \mat{A}_1 \hat{\mat{K}}(\infty)^{-1}
                \end{array}
	\right.
\end{equation}
where $\lambda_1$ is the user-chosen point and we are solving for unknowns $\mat{A}_1$ and $\mat{B}_1$.
Any nonlinear solver can be employed to solve Eq.~\eqref{eq:app2}; we used the default trust-region algorithm available with MATLAB.

Our previous work used the first approach since the correlation matrices $\mat{G}$ and $\mat{H}$ were defined differently, allowing access to higher-order derivatives at $t=0$.  
In contrast, for the new correlation matrices defined in Eq.~\eqref{eqn:gh}, high-order information is no longer available.  
As such, we use the second approach in this current work.
We note that the choice of the interpolation points is somewhat \textit{ad hoc}: for the present study, we choose points to capture the 
peak and asymptotic values of $\hat{\mat{K}}(\lambda)$.  
 As shown in Figure~\ref{f:theta}, we see a pronounced peak in $\hat{\mat{K}}(\lambda)$ which is indicative of dynamics in the \ac{BnBr} time domain fluctuating more prominently toward the origin (for more on the relationship between $\mat{K}(t)$ and $\hat{\mat{K}}(\lambda)$, see \cite{Davies2010}).
Thus, we chose points close to $\lambda = 0 $ so that we could more accurately capture this behavior. 
This choice of points does affect the quality of the approximation; an example is shown in Section~\ref{s:int_pts}.
The optimal choice of the $\lambda$ requires further investigation.

\subsection{Representing the \ac{GLE} with extended dynamics driven by white noise}
\label{sec:ex_dynamics}
Once we have determined coefficients of the rational memory term, we can construct a new approximate \ac{GLE} system~\cite{Lei2016}.
We illustrate this construction by deriving the extended system in the first-order case.  
We know from Eq.~\eqref{e:ratK} that the first-order rational approximation for $\hat{\mat{K}}$ is given by
\begin{equation}
    \hat{\mat{K}}(\lambda) \approx (\mat{I} - \mat{B}_1\lambda)^{-1}(\mat{A}_1\lambda).
    \nonumber
\end{equation}
Taking the inverse Laplace transform, $\mathcal{L}^{-1}$, we obtain
\begin{equation}
    \mat{K}(t) = \mathcal{L}^{-1}\left\{\hat{\mat{K}}(\lambda)\right\} \approx \mat{A}_1 e^{\mat{B}_1 t}.
    \nonumber
\end{equation}
%
Let us define the auxiliary variable $\vec{d}(t): [0, \infty) \mapsto \mathbb{R}^{MN}$, where $M$ is the order of the rational approximation, as
\begin{equation}
    \vec{d}(t) = -\int_0^t \mat{K}(t-\tau) \dot{\vec{q}}(\tau)d\tau + \vec{R}(t) \nonumber.
\end{equation}
With $M=1$ for this derivation, we next define  $\vec{d}_1(t)$, the auxiliary variable in the first-order case, as
\begin{equation}
	\vec{d}_1(t) = -\int_0^t \mat{A}_1 e^{\mat{B}_1 (t-\tau)} \dot{\vec{q}}(\tau)d\tau + \vec{R}(t).
\end{equation}
Using the Leibniz integral rule, we can differentiate $\vec{d}_1(t)$:
\begin{equation}
    \dot{\vec{d}}_1(t) = - \mat{A}_1 \dot{\vec{q}}(t) - \mat{B}_1 \int_0^t \mat{A}_1 e^{\mat{B}_1 (t-\tau)} \dot{\vec{q}}(\tau)d\tau + \dot{\vec{R}}(t).
    \nonumber
\end{equation} 

 Note that $\vec{R}(t)$ is assumed to be colored noise and not white noise, and as such is differentiable.
As discussed in Section~\ref{sec:fdt}, $\vec{R}(t)$ obeys the \ac{FDT}. 
Deferring the details to Section~\ref{sec:fdt}, the detailed {colored} noise can be expressed in terms of the initial condition $\vec{d}_1(0)$ and a simple {white} noise term  $\vec{W}_1(t)$:
\begin{equation}
    \vec{R}(t) = \int_0^t e^{\mat{B}_1 (t-\tau)} \vec{W}_1(\tau)d\tau + e^{\mat{B}_1 t}\vec{d}_1(0),
    \nonumber
\end{equation}
which is further discussed in the next section.
We remark that $\mat{B}_1$ is negative and the first term of $\vec{R}(t)$ is well-behaved for large $t$. 
Using the Leibniz integral rule again, we can write $\dot{\vec{d}}_1(t)$:
\begin{align}
    \dot{\vec{d}_1}(t) =& - \mat{A}_1 \dot{\vec{q}}(t) - \mat{B}_1 \int_0^t \mat{A}_1 e^{\mat{B}_1 (t-\tau)} \dot{\vec{q}}(\tau)d\tau \nonumber \\
    & +\vec{W}_1(t) + \mat{B}_1 \int_0^t e^{\mat{B}_1 (t-\tau)} \vec{W}_1(\tau) d\tau + \mat{B}_1 e^{\mat{B}_1 t}\vec{d}_1(0).
    \label{eq:extra1}
\end{align} 
Note that
\begin{align}
    \mat{B}_1 \int_0^t e^{\mat{B}_1 (t-\tau)} \vec{W}_1(\tau) d\tau + \mat{B}_1 e^{\mat{B}_1 t}\vec{d}_1(0) &= \mat{B}_1 \vec{R}(t) \nonumber \\
    \mat{B}_1 \int_0^t \mat{A}_1 e^{\mat{B}_1 (t-\tau)} \dot{\vec{q}}(\tau)d\tau &= \mat{B}_1 \vec{d}_1(t) - \mat{B}_1 \vec{R}(t) \nonumber
\end{align}
and so Eq.~\eqref{eq:extra1} can be written as
\begin{equation}
    \dot{\vec{d}_1}(t) = \mat{B}_1 \vec{d}_1(t) - \mat{A}_1 \dot{\vec{q}}(t) + \vec{W}_1(t) \nonumber
\end{equation}
to obtain the first-order approximate \ac{GLE} system:
\begin{eqnarray}
    \dot{\vec q} &=& \mat{M}^{-1}\vec{p}, \nonumber \\
    \dot{\vec p} &=& \vec{F(q)}+\vec{d}_1, \\
    \dot{\vec d}_1 &=& \mat{B}_1\vec{d}_1 - \mat{A}_1 \dot{\vec{q}}+\vec{W}_1. \nonumber
\end{eqnarray}
Higher-order approximations are obtained by generalizing the procedure above to obtain
\begin{eqnarray}
    \dot{\vec q} &=& \mat{M}^{-1}\vec{p}, \nonumber \\
    \dot{\vec p} &=& \vec{F(q)}+\mat{Z}^T\vec{d}_M, \\
    \dot{\vec d}_M &=& \mat{B}\vec{d}_M - \mat{Q}\mat{Z}\dot{\vec{q}} +\vec{W}_M, \nonumber
\end{eqnarray}
where the symmetric positive definite matrix $\mat{Q} \in \mathbb{R}^{MN \times MN}$ and the matrix 
$\mat{Z} \in \mathbb{R}^{MN \times N}$ are determined by matching Eq.~\eqref{e:ratK} with 
the equation $\mat{K}(t) = \mat{Z}^T e^{\mat{B}t} \mat{Q} \mat{Z}$.
$\mat{Q}$ is the covariance matrix of the auxiliary vector $\vec{d}_{M}$ under equilibrium.
The matrix $\mat{B}$ is dependent on order; e.g., a fourth-order approximation would have the form
\begin{equation}
    \mat{B} = 
    \begin{pmatrix} 0 & 0 & 0 & \mat{B}_4\\
        \mat{I} & 0 & 0 & \mat{B}_3\\
        0 & \mat{I} & 0 & \mat{B}_2\\
        0 & 0 & \mat{I} & \mat{B}_1
    \end{pmatrix}.
    \nonumber 
\end{equation}

\subsection{Initial and noise conditions to satisfy the second \ac{FDT}}\label{sec:fdt}
Recall that $\vec{R}(t)$ in Eq.~\eqref{e:md} simulates system noise as a colored noise that must satisfy the second \ac{FDT}.
Through our extended \ac{GLE} system, we can replace $\vec{R}(t)$ with a simpler white noise term $\vec{W}(t)$ and choose the initial and noise 
conditions for $\vec{W}(t)$ and $\vec{d}(t)$ to ensure that the colored noise generated by these extended dynamics also satisfies the second \ac{FDT}~\cite{Lei2016}.
For the first-order approximation, the initial and noise conditions are 
\begin{equation}
    \begin{aligned}
        \left\langle\vec{d}_1(0)\vec{d}_1(0)^T\right\rangle &= \beta^{-1}\mat{A}_1 \\
        \left\langle\vec{W}_1(t)\vec{W}_1(t')^T\right\rangle &= -\beta^{-1}\left(\mat{B}_1 \mat{A}_1 +\mat{A}_1 \mat{B}_1^T\right) \delta(t-t'),
    \end{aligned}
    \label{e:noise_1}
\end{equation}
and for higher-order approximations, the initial and noise conditions are
\begin{equation}
    \begin{aligned}
        \left\langle\vec{d}_M(0)\vec{d}_M(0)^T\right\rangle &= \beta^{-1}\mat{Q} \\
        \left\langle\vec{W}_M(t)\vec{W}_M(t')^T\right\rangle &= -\beta^{-1}\left(\mat{B}\mat{Q} + \mat{Q}\mat{B}^T\right)\delta(t-t') 
    \end{aligned}
    \label{e:noise}
\end{equation}
For other work on \acp{GLE} for systems driven by white noise, see~\cite{Hudson2018} and~\cite{Zhu2020a}.

The extended \ac{GLE} system approximations eliminate costly integration of the exact memory term, which depends on the system history, by replacing it with an extended system of stochastic differential equations.
The accuracy of this approximation improves with increasing order which involves reformulation of the matrix $\mat{B}$ and recalculation of the matrices $\mat{Q}$ and $\mat{Z}$.
These computations are relatively simple to perform for the low-order approximations needed to model small molecular systems (see Section~\ref{s:results} of this manuscript).
Overall, our method provides substantial dimension-reduction and results in significant increase in computational tractability.

\subsection{Simulation setup}\label{sec:sim_setup}
\begin{figure}
    \centering
    \includegraphics[width=0.2\linewidth]{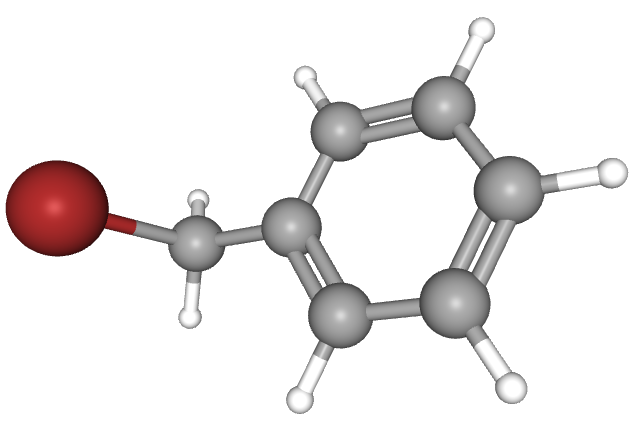} 
    \caption{The \acf*{BnBr} molecule with carbon atoms shown in gray, hydrogen in white, and bromine in red.}\label{f:blb}
\end{figure}
Simulations were run using GROMACS~\cite{Berendsen1995} and the general AMBER force field~\cite{Wang2004}.
We performed 360 simulations of a single \ac{BnBr} molecule (Figure~\ref{f:blb}), which is comprised of 15 atoms, in a solvent consisting of 1011 water molecules in a $\left( 3.14216~\text{nm} \right)^3$ domain.
We used a \ac{NVT} ensemble with a Nos\'{e}-Hoover thermostat~\cite{Hoover1985} at 300 K.
Each simulation ran for 10 ns with a time step of 2 fs.
The particle-mesh Ewald method~\cite{Darden1993} was used for long-range electrostatics. 
All \ac{BnBr} bond lengths were constrained using the LINCS algorithm~\cite{Hess1997}; we note that this significantly reduces the dimension of the molecular conformational space.
The \ac{BnBr} positions were stored at every time step.

\begin{figure}
    \centering
    \includegraphics[width=0.4\linewidth]{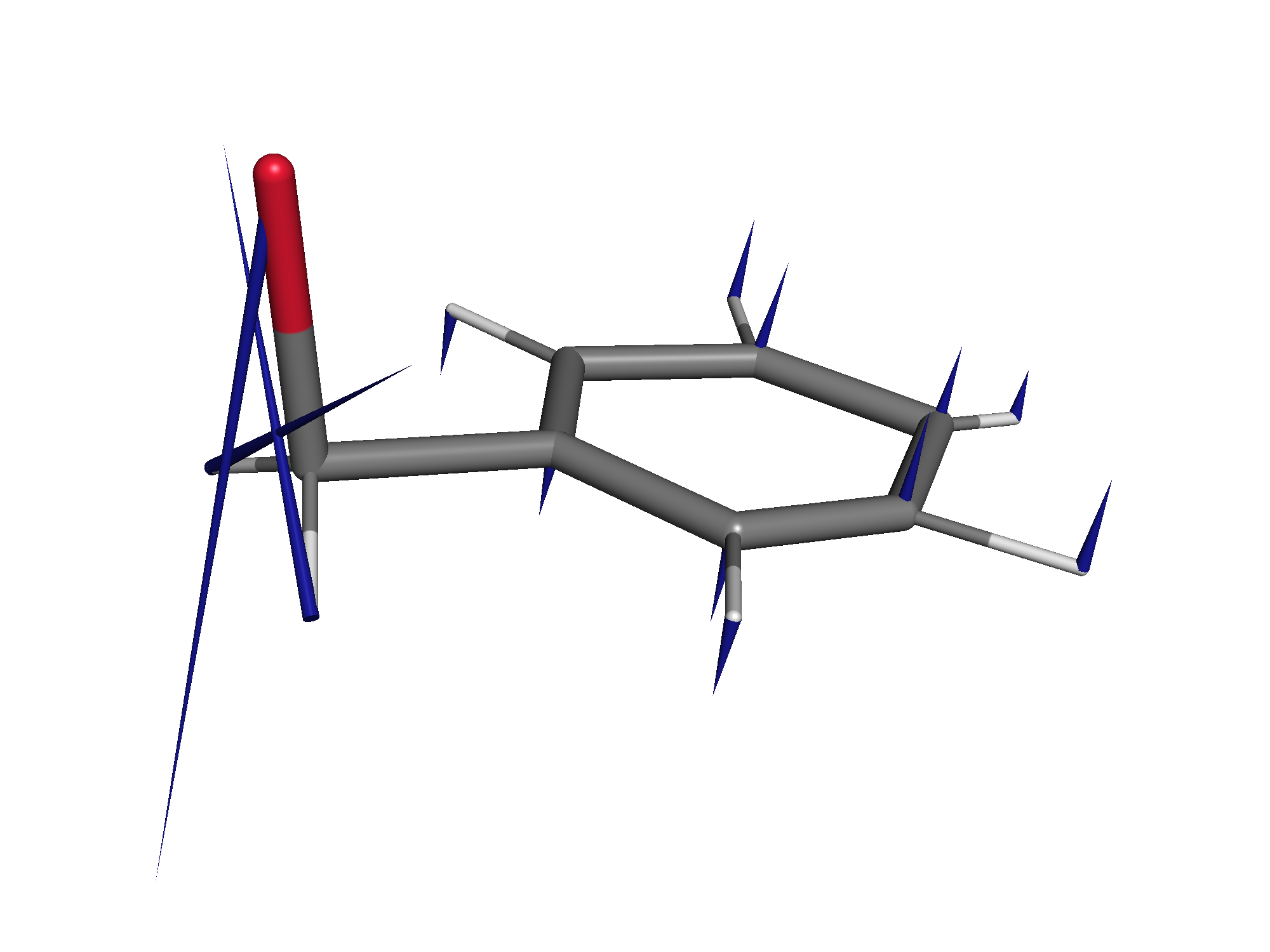} 
    \caption{Porcupine plot of first eigenvector showing dominant motion of \acf{BnBr}.}
    \label{f:porcupine}
\end{figure}

As a post-processing step, translational and rotational degrees of freedom were removed from the trajectory using the GROMACS function \emph{trajconv}.
 Averaging was then done over the 360 total trajectories.
We performed \ac{PCA} on these trajectories and checked for convergence by splitting the post-processed trajectories into equal halves and calculating the \acp{PDF} of each half for the first few principal components.
These \acp{PDF} of both halves matched well with each other, indicating the simulation had converged.
As the first principal component accounted for 63\% of the observed variance, we found it sufficient to use this as our single \ac{CV} for the purposes of illustration in this paper.
To physically interpret and visualize the results of \ac{PCA}, we can generate a porcupine plot showing the motion along an eigenvector.
In particular, the porcupine plot of the first eigenvector, Figure~\ref{f:porcupine}, shows dominant motions of \ac{BnBr}, with the direction and length of each ``quill" showing the direction and magnitude of motion, respectively.
In particular, we can see the bromomethyl group contributes prominently to the motion of the first mode.

From this, we constructed the correlation matrices and solved for the unknown rational coefficients for zeroth- to fourth-order approximations as described above.
Note that a zero-order approximation is simply a Markovian approximation, with the integral term in Eq.~\eqref{e:md} simplifying to  $\hat{\mat{K}}(\infty)\dot{\vec q}(t)$; see Ma et al.\ for details~\cite{Ma2016}.

\section{Results}\label{s:results}

The following section presents results testing our approximation against exact methods.
In particular, we assess comparisons using the memory kernel, position autocorrelation, velocity autocorrelation, and mean first-passage time.

\subsection{Memory kernel}

The \ac{PDF} $\rho(\vec{q})$ of the \ac{CV} defined in Section~\ref{sec:preliminary} can be calculated using kernel density estimation on samples from the \ac{MD} trajectory. 
This \ac{PDF} can be used to calculate the free energy $U(\vec{q})$ 
\begin{equation}
    U(\vec{q}) = -\beta^{-1} \ln \left( \rho(\vec{q}) \right)
    \label{eq:pmf}
\end{equation}
which, in turn, can be used to calculate the mean force
\begin{equation}
    \vec{F}(\vec{q}) = -\nabla U(\vec{q}).
\end{equation}
With $\vec{F}(\vec{q})$, we are able to sample $\mat{G}(t)$ and $\mat{H}(t)$ and construct the Laplace transform of the memory term $\hat{\mat{K}}(\lambda)$ based on the numerical approach introduced in Section~\ref{sec:memory_approx}.

\begin{figure}
    \centering
    \includegraphics[width=0.5\linewidth]{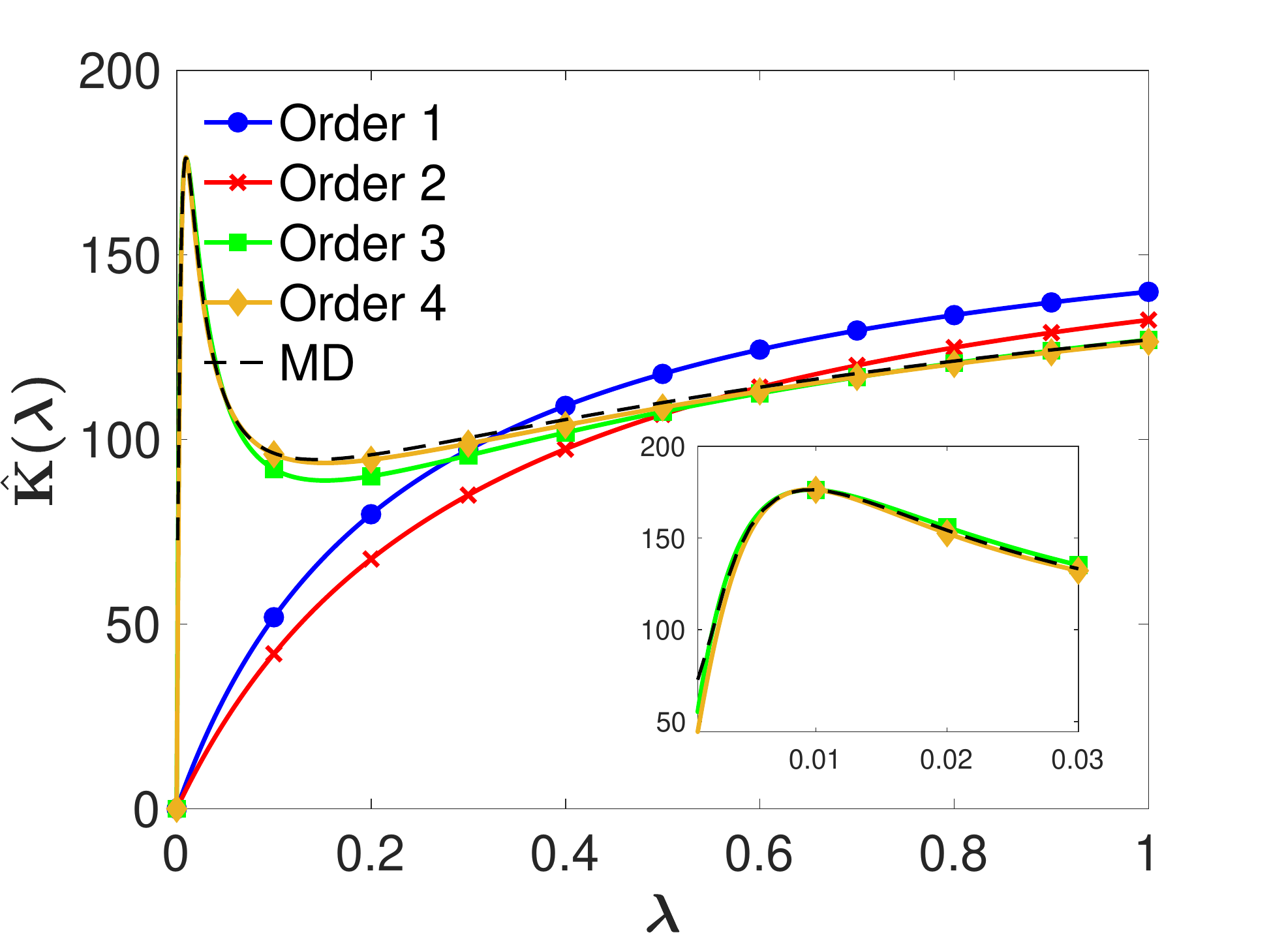} 
    \caption{Memory kernel in Laplace space from \ac{MD} simulation versus kernels constructed using data-driven \ac{GLE} approximations of varying orders.
    Inset: Close-up of third- and fourth-order approximations capturing the pronounced peak of $\hat{\mat{K}}(\lambda)$.}\label{f:theta}
\end{figure}

The exact $\hat{\mat{K}}(\lambda)$  calculated from our simulations shows a pronounced peak near $\lambda = 0.01$.
Since this peak is indicative of oscillations in $\mat{K}(t)$ (i.e.~oscillations back in the time domain), good approximation 
of this peak is important for capturing system dynamics.
As shown in Figure~\ref{f:theta}, the first- and second-order approximations do not reproduce the peak; however, the third- and fourth-order rational functions have enough interpolation points for an accurate model.

\subsection{Autocorrelation functions}
\begin{figure}
    A.~\includegraphics[width=0.45\linewidth]{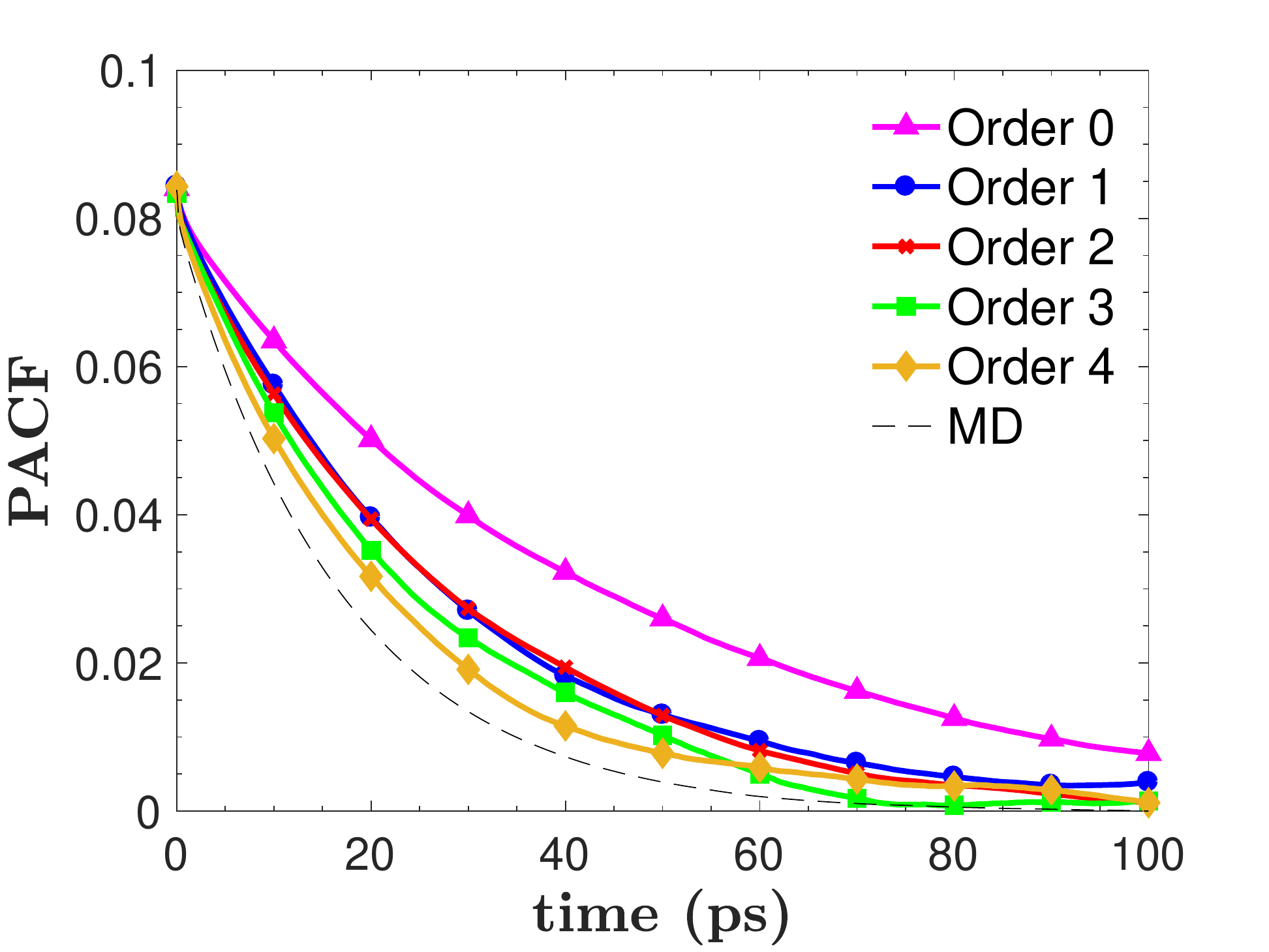} 
    B.~\includegraphics[width=0.45\linewidth]{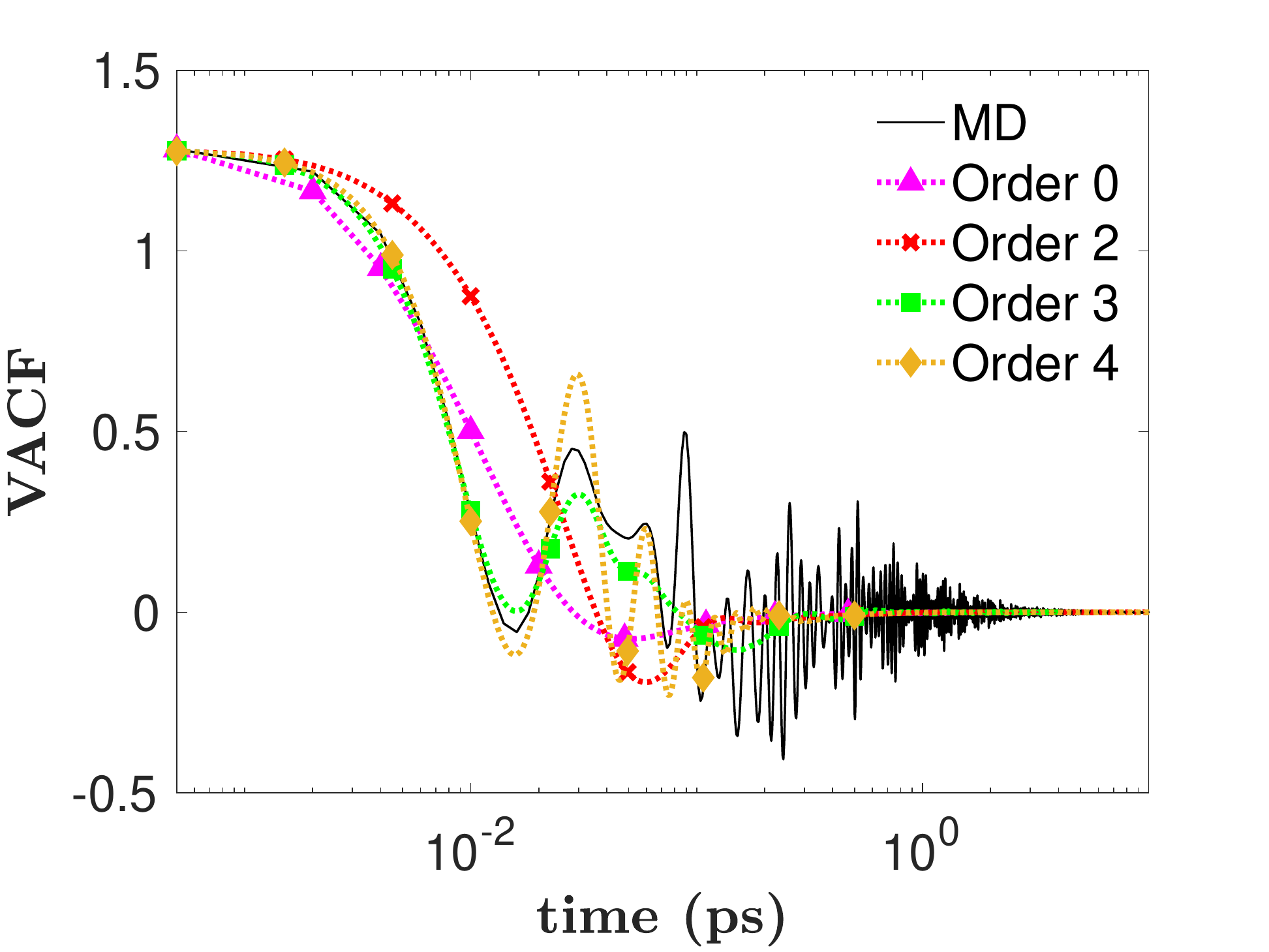} 
    \caption{\ac*{PACF} (A) and \ac*{VACF} (B) for exact \ac{MD} data compared to approximate \ac{GLE} simulations.}\label{f:acf}
\end{figure}

The \ac{PDF} tests the equilibrium properties of the approximation.
To test dynamic properties, we computed both the \ac{PACF} $\langle\vec{q}(t)\vec{q}(0)^T\rangle$ as well as the \ac{VACF} $\langle\dot{\vec{q}}(t)\dot{\vec{q}}(0)^T\rangle$ 
and compared the resulting approximate trajectories to data calculated directly from the original \ac{MD} simulation.
The results are shown in Figure~\ref{f:acf}.
The accuracy of the \ac{PACF} increases with increasing order of the \ac{GLE} approximation, with all performing better than the zero-order Markovian approximation.
Likewise, the accuracy of the \ac{VACF} also increases with increasing order of the \ac{GLE} approximation.
Long-time oscillations occur in the \ac{VACF} of \ac{BnBr}; we found such behavior is due to the following:
\begin{itemize}
	\item the strong intramolecular covalent bond interactions and the dominant motions of BnBr. 
	\item  the number of peaks in the FFT of the velocity autocorrelation is of comparable order to the number of peaks in the vibrational spectrum of BnBr.
\end{itemize}
These oscillations make it particularly challenging to approximate; inaccuracies in these autocorrelations may lead to misinterpretation of the underlying nature of the system dynamics.
The third- and fourth-order approximations reproduce both the \ac{PACF} and \ac{VACF} fairly well.
Recall that we applied LINCS constraints to all \ac{BnBr} bond lengths, which reduced the amount of noise in the \ac{VACF} of the principal components, and likely allowed for easier approximation of the \ac{GLE} terms.

\subsubsection{Selecting interpolation points}\label{s:int_pts}

\begin{figure}
    \centering
    \includegraphics[width=0.5\linewidth]{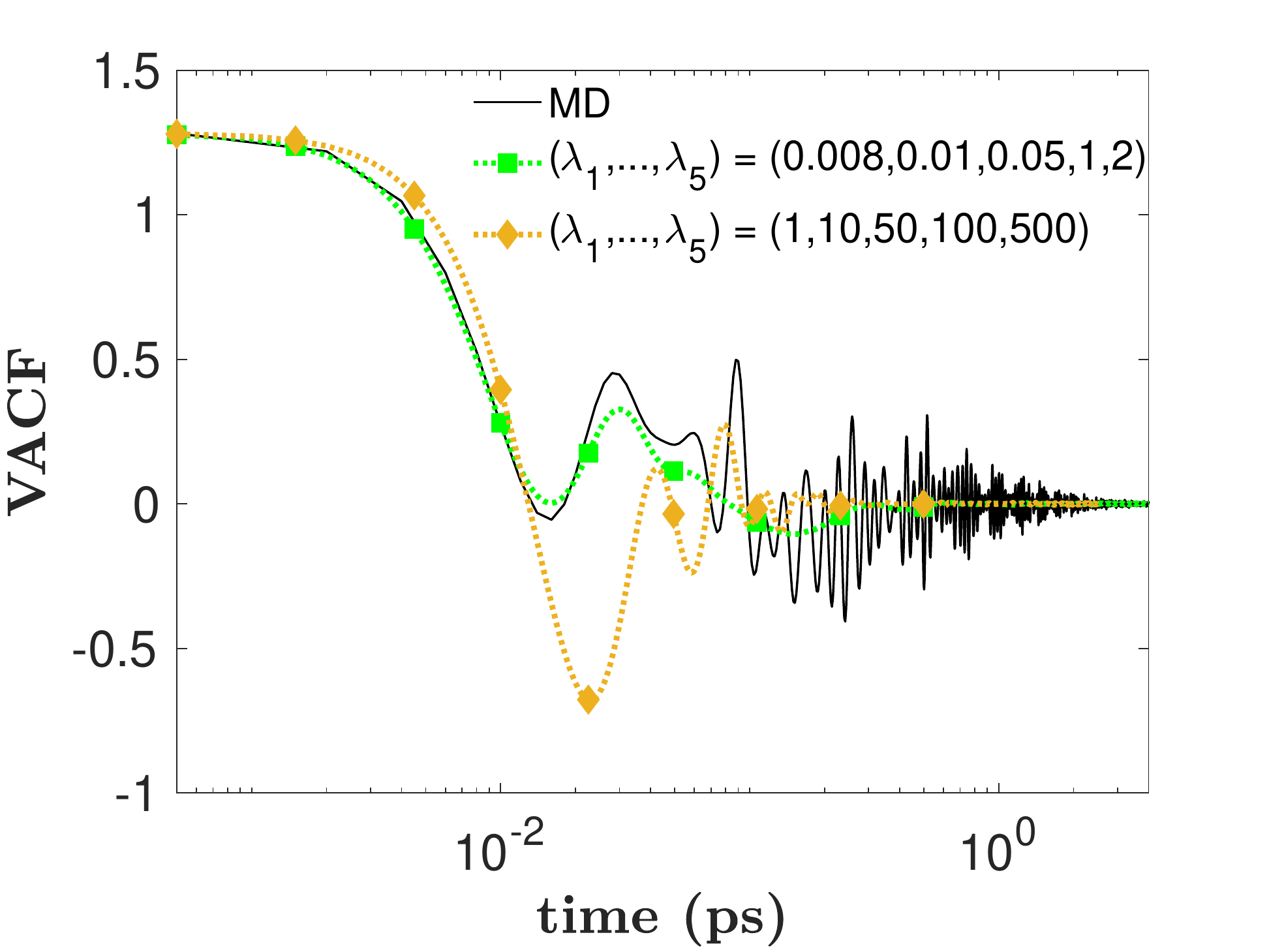} 
    \caption{Third-order \ac{VACF} approximation using two different sets of interpolation points $\lambda_i$, $i=1, \ldots, 5$.}\label{f:diff_interp}
\end{figure}

Recall that we construct an order-$M$ rational memory term through regression with user-selected values $\lambda_1, \lambda_2, \ldots, \lambda_{2M-1} \in (0,\infty)$.
Figure~\ref{f:diff_interp} compares the \acp{VACF} of two third-order approximations constructed using two different sets of interpolation points.
We see that the approximation using shorter-time interpolation points more accurately reproduces the \ac{VACF}.
As shown in Figure~\ref{f:theta}, this increase in accuracy is due to the regression model sufficiently capturing the pronounced peak in $\hat{\mat{K}}(\lambda)$, which occurs close to $\lambda=0$.

\subsection{Mean first-passage time}
\begin{figure}
    \centering
    A.~\includegraphics[width=0.45\linewidth]{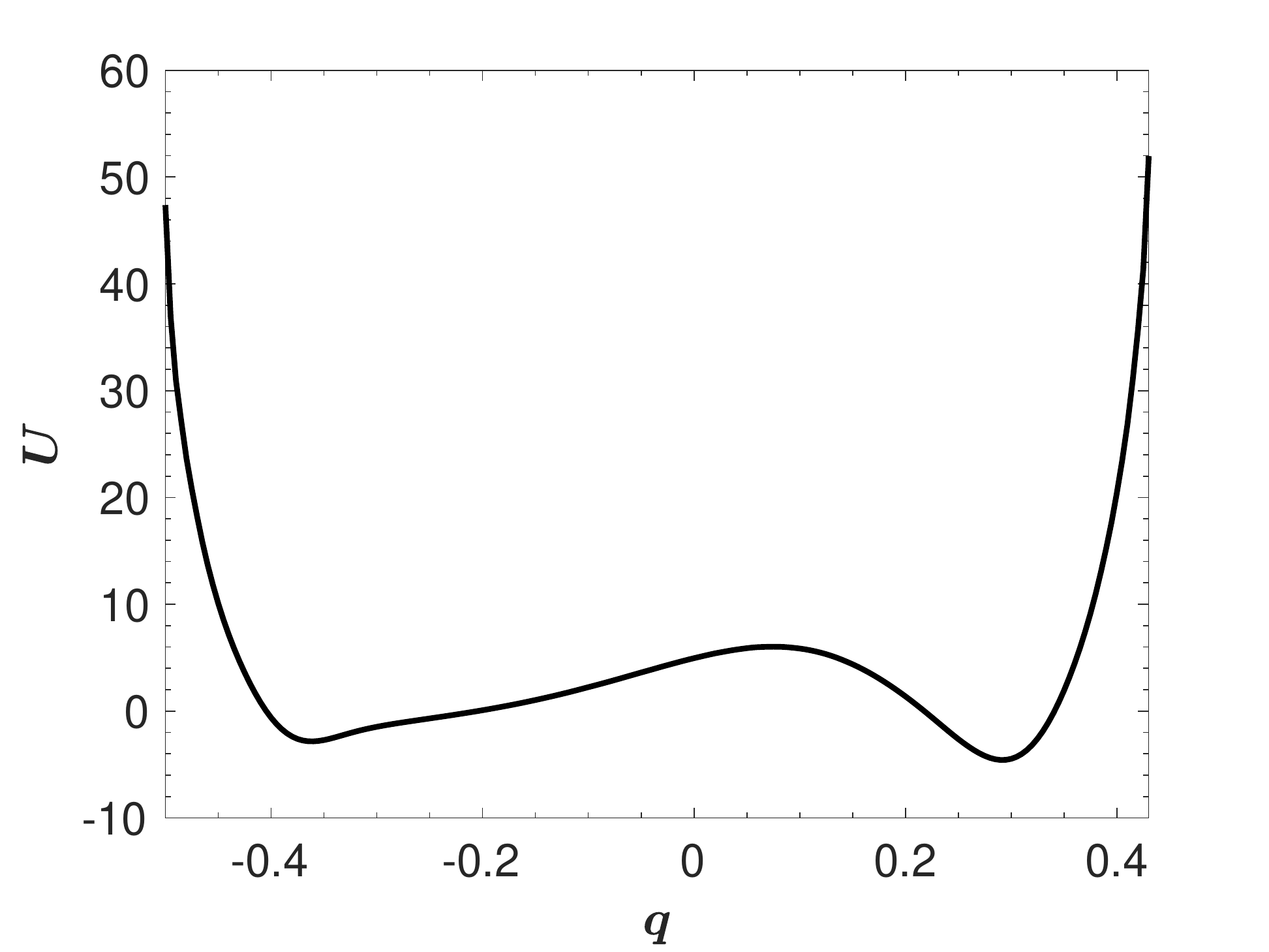} 
    B.~\includegraphics[width=0.45\linewidth]{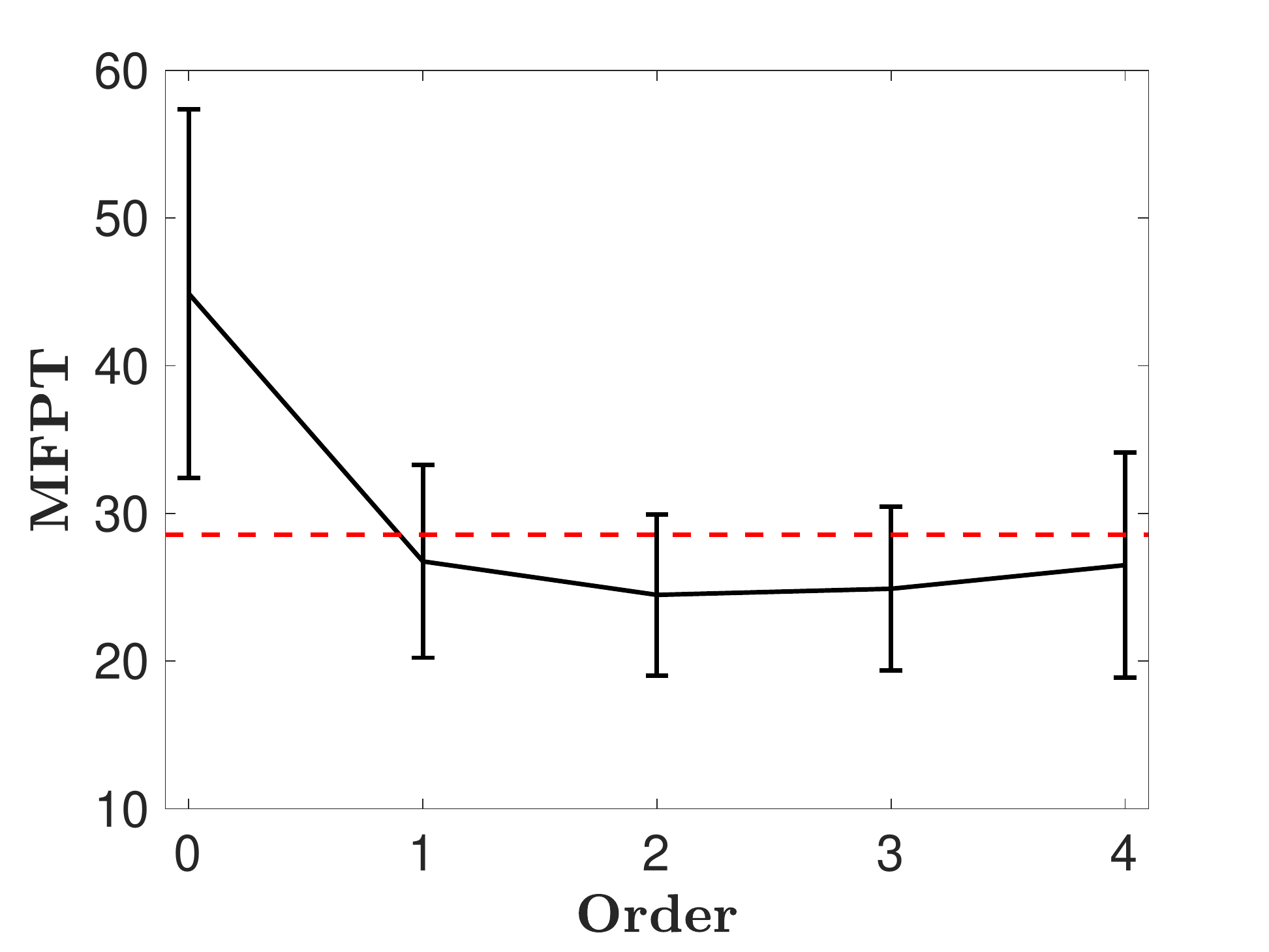}
    \caption{Mean first-passage time between states in a double-well potential of mean force.
    A.~Double-well potential $U\vec{(q)}$ calculated from the \ac{PDF} $\rho (\vec{q})$.
    B.~\ac{MFPT} (ps) from the approximate \ac{GLE} compared to the exact \ac{MD} data (red dotted line), shown with a 95\% confidence interval.}\label{f:well}
\end{figure}

Predicting non-equilibrium properties such as \ac{MFPT} between states is a challenging test for the \ac{GLE} approximation 
since this statistic from the original \ac{BnBr} \ac{MD} simulations was not known or used \emph{a priori} in the data-driven parametrization of the \ac{GLE}.
From the density $\rho$, we can calculate the potential of our first principal component $U\vec{(q)} = -\beta^{-1} \ln(\rho)$ which has two wells, as shown in Figure~\ref{f:well}A.
Denoting the left potential well as state ``A'' and the right as state ``B'', we define the \ac{MFPT} as the mean time for a particle starting at an initial state to cross the peak maximum into the other state.  
In this example, this maximum occurs at $q = 0.075$; thus state A is defined as $q < 0.075$ and state B is defined as $q > 0.075 $.
Figure~\ref{f:well}B shows a comparison for all orders of the \ac{MFPT} and the exact \ac{MD} trajectory.
The Markovian approximation fails to accurately reproduce the \ac{MFPT}, while the higher-order \ac{GLE} approximations show significantly better agreement with the \ac{MD} results.

\section{Conclusion}\label{s:conclusion}

As full \ac{MD} simulations often require intensive computational resources and long run times to achieve ergodicity, researchers have increasingly relied on reduced-order modeling for simulation.  
In particular, the \ac{GLE} has seen a resurgence in popularity, as it provides a convenient description of coarse-grained dynamics.
While the exact \ac{GLE} can significantly reduce problem size and difficulty, the memory kernel of the \ac{GLE} relies on past-system history and is often hard to characterize and compute.
To mitigate this, previous work introduced a data-driven approximation to the \ac{GLE}.
While there is cost associated with sampling from the exact system, it can be computationally cheaper than running the full MD simulation for very long times.
Directly sampling from correlation functions of exact system dynamics, we replaced the memory kernel with a rational approximation and carefully introduced an auxiliary variable and white-noise term to convert the \ac{GLE} into an extended system that does not rely on past-system history.
Additionally, accuracy is adaptively affected by the chosen order of the rational approximation.
This current work extends a data-driven approximation of the \ac{GLE}~\cite{Lei2016} to more complex and realistic molecules.
Using \ac{BnBr} as our test case, our comparison of exact \ac{MD} simulation against the approximation shows observables are reproduced well using relatively low orders for the rational term.

There are multiple avenues for future work that further develops modeling capability of complex systems.
While we were able to represent \ac{BnBr} system dynamics with a single \ac{CV}, accurate construction 
of the memory term and reduced dynamics remains a challenging
task as the complexity of the system increases, since the condition number of  $\mat{B}$ may become large.
Thus, it would be necessary to test for robustness on systems where the \ac{CV} dimension is higher than one, as is done in~\cite{Lee2019}.
Towards this end, to alleviate such difficulties, we are developing a regularization approach for the \ac{GLE} approximation by
formulating the memory kernel construction as an optimization problem.
Furthermore, while we were able to use an unbiased density to compute the force term in this current work, it is more difficult 
to compute this term with respect to higher-dimensional spaces.
To ensure adequate sampling of the energy surface, enhanced sampling methods may need to be paired with the 
data-driven \ac{GLE} approximation in order to give robust results.

\section*{Acknowledgments}

We thank Peiyuan Gao for helpful discussions.
This work was performed using resources through Research Computing at Pacific Northwest National Laboratory.
FG acknowledges support from the \ac{DOE} Office of \ac{ASCR} through the \ac{ASCR} Distinguished Computational Mathematics Postdoc Project under ASCR Project 71268.
HL and NAB acknowledge support from NIH grant GM069702.


\bibliographystyle{unsrtnat}
\biboptions{sort&compress,numbers}
\bibliography{Grogan_JCP_manuscript_final}

\appendix

\section*{Acronyms}

\begin{acronym}
    \acro{ASCR}{Advanced Scientific Computing Research}
    \acro{BnBr}{benzyl bromide}
    \acro{CV}{collective variable}
    \acro{DOE}{Department of Energy}
    \acro{FDT}{fluctuation-dissipation theorem}
    \acro{GLE}{generalized Langevin equation}
    \acro{MFPT}{mean first-passage time}
    \acro{MD}{molecular dynamics}
    \acro{NVT}{constant number-volume-temperature}
    \acro{PACF}{position autocorrelation function}
    \acro{PCA}{principal component analysis}
    \acro{PDF}{probability density function}
    \acro{VACF}{velocity autocorrelation function}
\end{acronym}

\end{document}